\documentclass[twocolumn, secnumarabic, amssymb, amsmath,nobibnotes, aps, prl, showpacs]{revtex4}
\begin{document}
\title{Mean Entropies}
\author{B. H. Lavenda}
\email{bernard.lavenda@unicam.it}
\affiliation{Universit\'a degli Studi, Camerino 62032 (MC) Italy}
\date{\today}
\newcommand{\sumi}{\sum_{i=1}^{N}\,}

\pacs{05.70Ln,05.20.Gg,05.40.-a}
\begin{abstract}
Entropies must correspond to mean values for them to be measurable. The Shannon entropy  corresponds to the weighted arithmetic mean, whereas  the R\'enyi entropy  corresponds to the exponential mean. These means refer to code lengths, which are converted into entropies by replacing the length of a sequence by the negative logarithm of the probability of its occurrence. Only affine and exponential generating functions of means  preserve the property of additivity and invariance  under translations, and hence are Kolmogorov-Nagumo functions,  resulting in the Shannon and R\'enyi entropies, respectively.  Pseudo-additive entropies are generating functions of  means of order $0\le\tau<1$, which is the exponential of the R\'enyi entropy, or in the $\tau=0$ limit, the Shannon entropy. Means of any  order  cannot be expressed as escort averages because such averages contradict the fact that the means are monotonically increasing functions of their order.   Exponential mean error functions of R\'enyi, in general, and Shannon, in particular, are shown to be measures of the extent of a distribution.
\end{abstract}
\maketitle
\section{Entropy and coding theory}
In his seminal paper on entropy and information, R\'enyi \cite{Renyi} laid down the fundamental properties of entropy and its relation to a measure of information. Many of these tenets have been all but abandoned \cite{T}.\par
Basing himself on the relation between the mean code length,
\begin{equation}
L_a(n)=\sumi p_in_i \label{eq:L-arith}
\end{equation}
and the Shannon entropy, R\'enyi argued convincingly that any putative candidate for an entropy should be a mean. Here, $p=(p(x_1),p(x_2),\ldots,p(x_N))$ is the set of probabilities of $N$ input symbols, $x=(x_1,x_2,\ldots,x_N)$, that are to be encoded using an alphabet of size $D$. Each $p(x_i)>0$, and the distribution is complete, $\sumi p(x_i)=1$. The $x_i$ represent a sequence of $n_i$ characters taken from the alphabet.\par  There exists a uniquely decipherable code with lengths $n_i$ iff
\begin{equation}
\sumi D^{-n}\le 1, \label{eq:Kraft}
\end{equation}
which is known as Kraft's inequality \cite{Feinstein}. The equality is automatically guaranteed by setting
\begin{equation}
n_i=-\log_D p(x_i), \label{eq:n}
\end{equation}
which represents the amount of information received by knowing that an event of probability $p(x_i)$ has occurred.\par Introducing (\ref{eq:n}) into the weighted arithmetic average (\ref{eq:L-arith}) gives Shannon's expression for the entropy,
\begin{equation}
S_S\left(p\right)=-\sumi p_i\log_Dp_i, \label{eq:Shannon}
\end{equation}
 where the abbreviation $p_i=p(x_i)$ has been introduced. R\'enyi addressed the question of what other entropies are obtainable when the weighted arithmetic mean, (\ref{eq:Shannon}), is replaced by a generalized mean \cite{HLP}
\begin{equation}
S(n)=\phi^{-1}\left\{\sumi p_i\phi(n_i)\right\}=
\mathfrak{M}_{\phi}(n), \label{eq:mean}
\end{equation}
where $\phi(x)$ is a strictly monotonic and continuous function that possesses an inverse $\phi^{-1}(x)$.\par
The generating function $\phi$ must be so chosen that the generalized mean (\ref{eq:mean}) possess the following properties. First and foremost, it must have the property of  \emph{additivity\/}.  If $q=(q(x_1),q(x_2),\ldots,q(x_N))$ represents another finite discrete probability distribution then the entropy of their direct product should be additive
\begin{equation}
S\left(p\otimes q\right)=S\left(p\right)+S\left(q\right).\label{eq:add}
\end{equation}
This is guaranteed by the fact that the events are independent so that their probabilities multiply, and the entropy satisfies the functional equation, (\ref{eq:add}). According to (\ref{eq:n}) multiplicative probabilities means that code lengths are additive, as they should be.\par Secondly, the entropy should possess the \emph{mean-value property\/}, which says that the entropy of the union of two distributions is the weighted arithmetic mean of the individual entropies.\par
Other properties listed by R\'enyi \cite{Renyi} that an entropy should have were symmetry, continuity, and normalization. 
\section{Equivalent means} In addition to these properties, any \emph{classical\/} expression for the entropy should be \emph{translation invariant\/}, which says that only entropy differences are measurable. For $N=2$, the generalized mean must satisfy
\begin{eqnarray}
\lefteqn{\phi^{-1}\left[p_1\phi(x_1+a)+p_2\phi(x_2+a)\right]}\nonumber\\
& = & \phi^{-1}\left[p_1\phi(x_1)+p_2\phi(x_2)\right]+a, \label{eq:KN}
\end{eqnarray}
where $a$ is a constant. Such functions, $\phi$, were first investigated by Kolmogorov \cite{Kolmogorov} and Nagumo \cite{Nagumo}, and will be referred to as KN functions. The only known strictly monotonic increasing solutions of the functional equation (\ref{eq:KN}) are \cite{Aczel}
\[\phi(x)=ax+b\]
affine, and
\[\phi(x)=a D^{\tau x}+b,\]
exponential functions, where $a\neq0$ and $\tau\neq0$.
\par
The affine solution leads immediately to the Shannon entropy (\ref{eq:Shannon}) under condition (\ref{eq:n}), whereas the exponential solution leads one to consider exponential mean lengths \cite{Campbell65}
\begin{equation}
L_e(n)=\frac{1}{\tau}\log_D\sumi p_iD^{\tau n_i} \label{eq:L-exp}
\end{equation}
whose corresponding entropies are \cite{Renyi}
\begin{equation}
S_R\left(p\right)=\frac{1}{\tau}\log_D\sumi p_i^{1-\tau} \label{eq:Renyi}
\end{equation}
under condition (\ref{eq:n}). \par
It is imperative to emphasize that R\'enyi worked with the code lengths as the independent variables, and not with the number of different sequences of length $n_i$ \cite{Khinchin}:
\begin{equation}
\omega_i=D^{n_i}. \label{eq:omega}
\end{equation}
For if he had considered (\ref{eq:omega}) as the set of independent variables he would have obtained  means of order $\tau$,
\[
\mathfrak{M}_{\tau}(\omega)=\left(\sumi p_i\omega_i^{\tau}\right)^{1/\tau}, \]
leading to the expression
\begin{equation}
S\left(p\right)=\left(\sumi p_i^{1-\tau}\right)^{1/\tau}
\label{eq:S-exp}
\end{equation}
as the entropy, rather than as the \lq exponential entropy\rq\ \cite{Campbell64}. Such an entropy would not possess the property of additivity that its logarithm would restore.\par
\section{Pseudo-additivity}
The foregoing discussions provides  a basis for understanding the property of \lq pseudo-additivity\rq\ that certain entropies have been found to possess \cite{AD,Lavenda}.\par Consider the function
\begin{equation}
\phi(n_i)=\frac{D^{\tau n_i}-1}{\tau}\label{eq:log}
\end{equation}
which becomes affine in the limit $\tau\rightarrow0$. Since (\ref{eq:log}) is a KN function, its mean is equivalent to (\ref{eq:L-exp}), and, consequently, to the R\'enyi entropy, (\ref{eq:Renyi}), when condition (\ref{eq:n}) is imposed. However, were we to introduce (\ref{eq:n}) directly into (\ref{eq:log}),
\begin{equation}
\phi\left(1/p_i\right)=\frac{p_i^{-\tau}-1}{\tau} \label{eq:T}
\end{equation}
and then take its mean value, we would  obtain an exponential entropy,
\[
\phi^{-1}\left\{\sumi p_i\phi\left(1/p_i\right)\right\}
=\mathfrak{M}_{\tau}\left(1/p\right)=D^{S_R\left(p\right)}, \]
rather than the R\'enyi entropy itself.\par The weighted arithmetic mean of (\ref{eq:T}), 
\begin{equation}
S_{HC}\left(p\right)=\sumi p_i\phi\left(1/p_i\right)=
\frac{\sumi p_i^{1-\tau}-1}{\tau}, \label{eq:HCDT}
\end{equation}
is known as the Havrda-Charv\'at \cite{HC}-Dar\'oczy \cite{D}-Tsallis \cite{T} (HCDT) entropy. The  HCDT entropy, (\ref{eq:HCDT}), is pseudo-additive in that it is the solution of
\begin{equation}
S_{HC}\left(p\otimes q\right)=S\left(p\right)+S\left(q\right)+\tau S\left(p\right)S\left(q\right), 
\label{eq:pseudo}
\end{equation}
and not of (\ref{eq:add}). The KN function, whose mean is equivalent  to the mean of (\ref{eq:T}), 
\[\mathfrak{M}_{\tilde{\phi}}\left(1/p\right)=
\mathfrak{M}_{\phi}\left(1/p\right),\]
is
\[\tilde{\phi}\left(1/p_i\right)=\tau\phi\left(1/p_i\right)+1.\]
Its weighted arithmetic mean, 
\[
\tilde{S}_{HC}\left(p\right)=\tau S_{HC}\left(p\right)+1, \]
satisfies the functional equation
\begin{equation}
\tilde{S}_{HC}\left(p\otimes q\right)=\tilde{S}_{HC}
\left(p\right)\tilde{S}_{HC}\left(q\right),\label{eq:multi}
\end{equation}
implying a power law solution \cite[p. 39]{Aczel}, and the complete loss of additivity. Hence, on the basis of equivalent means we can transform pseudo-additivity, (\ref{eq:pseudo}), into a multiplicative relation, (\ref{eq:multi}), showing that neither relation has any thermodynamic meaning regarding the lack of extensivity.  \par
As (\ref{eq:S-exp}) clearly shows, it is the logarithm of the mean of the HCDT entropy that has physical meaning, and this is the R\'enyi entropy. If we had insisted on working with code lengths, and not with their probabilities, we would have obtained the R\'enyi entropy directly from introducing (\ref{eq:n}) into the mean code length (\ref{eq:L-exp}).\par
\section{Exponential mean entropy bounds on mean number of sequences}
According to Jensen's inequality for a convex function
\begin{equation}
\sumi p_iD^{\tau n_i}\ge D^{\tau\sumi p_in_i}. \label{eq:Jensen}
\end{equation}
Taking logarithms and considering $\tau>0$, there results
\begin{equation}
\frac{1}{\tau}\log_D\left(\sumi p_iD^{\tau n_i}\right)
\ge\sumi p_in_i, \label{eq:exp-arith}
\end{equation}
asserting that the exponential mean of parameter $\tau>0$ is never inferior to the weighted arithmetic mean. This is most easily shown for small $\tau$. Expanding the exponent and then the logarithm   in powers of $\tau$, we get to lowest order  \cite{Burrows}
\begin{eqnarray*}
\lefteqn{\frac{1}{\tau}\log_D\sumi p_i D^{\tau n_i}}\\
& = & \sum_i p_i n_i+\frac{\tau}{2}\left\{\sumi p_i n_i^2-\left(
\sumi p_i n_i\right)^2\right\}+\cdots
\end{eqnarray*}
The term is the curly brackets is always positive since it is the variance of $n_i$. We must also restrict $\tau<1$ in order to ensure that the R\'enyi entropy is concave.\par On the strength of (\ref{eq:n}), inequality (\ref{eq:exp-arith}) shows that the R\'enyi entropy is bounded from below by the Shannon entropy, and since the entropy is maximum for a uniform distribution, we have the following hierarchy:
\[S_S\le S_R\le S_H,\]
where $S_H=\log_D N$ is the Hartley entropy.\par
In terms of the number of different  sequences of length $n_i$, Jensen's inequality (\ref{eq:Jensen}) becomes 
\[
\mathfrak{M}_{\tau}(\omega)=
\left(\sumi p_i\omega_i^{\tau}\right)^{1/\tau}
\ge\prod_{i=1}^N\omega_i^{p_i}=\mathfrak{M}_0(\omega),\]
which says that means of order $\tau>0$ can never be inferior to the geometric mean \cite{BB}, the weighted arithmetic mean ($\tau=1$) being a particular case. The mean inequality for the same order and different argument \cite[p. 14]{HLP},
\begin{equation}
\mathfrak{M}_{\tau}(\omega)\ge\mathfrak{M}_{\tau}\left(1/p\right), \label{eq:major}
\end{equation}
follows from the fact that according to the Kraft inequality $\omega_i\ge p_i^{-1}$ for all $i$.
The equality in (\ref{eq:major}) holds when (\ref{eq:n}) is satisfied. The mean number of sequences is bounded from below by the exponential of the R\'enyi entropy. As a problem in majorization, we  say that $\omega$ majorizes $p^{-1}$, $\omega\succ p^{-1}$. \par
A similar, but not identical, result was found by Campbell \cite{Campbell65,Campbell66}, who used H\"older's inequality \cite[p. 24]{HLP}
\[\mathfrak{M}_{\tau/\alpha}(\omega)\mathfrak{M}_1\left(1/\omega p
\right)\ge\mathfrak{M}_{\tau}\left(1/p\right),
\]
where $\alpha=1-\tau$, and the Kraft inequality, (\ref{eq:Kraft}), to obtain
\begin{equation}
\mathfrak{M}_{\tau/\alpha}(\omega)\ge\mathfrak{M}_{\tau}\left(1/p\right). \label{eq:Campbell}
\end{equation}
The condition for the equality in (\ref{eq:Campbell}) is \emph{not\/} (\ref{eq:n}), but, rather,
\begin{equation}
\omega_i^{-1}=p_i^{\alpha}\big/\sumi p_i^{\alpha},
\label{eq:escort}
\end{equation}
which has been referred to as an \lq escort probability\rq \cite{Beck}. Whereas (\ref{eq:major}) implies (\ref{eq:Campbell}),  the converse is not true. In other words, inequality (\ref{eq:Campbell}) holds for $\alpha=1$, as (\ref{eq:major}) clearly shows, so that (\ref{eq:escort}) must also hold for $\alpha=1$, which is (\ref{eq:n}). In other words, (\ref{eq:n}) is sharper than (\ref{eq:escort}). Moreover, since inequality (\ref{eq:Campbell}) holds for $\alpha<1$, the same must be true in (\ref{eq:escort}). This condition has apparently gone  unappreciated \cite{Tsallis99}.\par
\section{\lq Escort\rq\ Averages}
Escort averaging has been used in variational formulations that maximize the pseudo-additive entropy (\ref{eq:HCDT}) with respect to escort expectations of thermodynamic constraints \cite{Tsallis99}. Pragmatically speaking,  it leads to analytic expressions for the variational equations, which would otherwise not exist. If escort averaging has any meaning at all, it must yield viable expressions for the means, and functions of the means.\par
Mean entropies are special cases of generalized means, where the variables, $-\log p_i$, and their weights, $p_i$, are not independent. Rather than considering means of order $\tau$, for which a demonstration that the mean is a monotonically increasing function of its order is given in \cite{Burrows}, we shall consider the exponential entropy (\ref{eq:S-exp}) and the R\'enyi entropy (\ref{eq:Renyi}).\par
 Differentiating
\[S^{\tau}\left(p\right)=\sumi p_i^{1-\tau},\]
with respect to $\tau$ yields
\begin{equation}S^{\tau}\left[\log_DS+\tau\frac{d\log_DS}{d\tau}
\right]=-\sumi p_i^{1-\tau}\log_Dp_i.\label{eq:first}
\end{equation}
At a stationary point, $d\log_DS/d\tau=0$, and 
\begin{equation}
S_R=\log_D S=-\sumi p_i^{\alpha}\log_D p_i\bigg/\sumi p_i^{\alpha}, \label{eq:stat}
\end{equation}
where again $\alpha=1-\tau$. As $\alpha\rightarrow1$ the entropy (\ref{eq:stat}) transforms into the Shannon entropy, (\ref{eq:Shannon}). Expression (\ref{eq:stat}) states that the logarithm of the exponential entropy (\ref{eq:S-exp}), which is the R\'enyi entropy, is the escort average of $-\log_D p_i$. The second derivative of (\ref{eq:first}), evaluated at the stationary point is
\begin{eqnarray*}
\lefteqn{\tau\frac{d^2\log_D S}{d\tau^2}}\\
& = & \frac{\sumi p^{\alpha}
\left(\log_D p_i\right)^2}{\sumi p_i^{\alpha}}
-\left(\frac{\sumi p_i^{\alpha}\log_D p_i}{\sumi p_i^{\alpha}}\right)^2>0,
\end{eqnarray*}
because the right-hand side is the variance of $-\log_D p_i$ under escort averaging. Hence, $d^2\log_D S/d\tau^2>0$ for $\tau>0$, and $d^2\log_D S/d\tau^2<0$ for $\tau<0$, implying that there are two extrema: a local maximum for $\tau<0$ and a local minimum for $\tau>0$. This requires  $d\log_DS/d\tau\le0$ at $\tau=0$, where the equality sign applies to the degenerate case where the extremes coincide in a point of inflection at $\tau=0$.\par Now, if it can be shown that $d\log_D S/d\tau>0$ at $\tau=0$, then $\log_D S$ has no extrema as a function of $\tau$, and, is, in fact, a monotonically increasing function of $\tau$ for all values of $\tau$. From this we will conclude that $\log_D S$ cannot be expressed as an escort average (\ref{eq:stat}), derived from a stationary condition, since such a condition does not exist.\par
Writing (\ref{eq:first}) in the form
\[S^{\tau}\frac{d\log_D S}{d\tau}=-
\frac{1}{\tau}\left\{\sumi p_i^{1-\tau}\log_D p_i+S^{\tau}\log_D S\right\},\]
it is apparent that the ratio on the right-hand side is of the form $0/0$ as $\tau\rightarrow0$ because $\log_D S\rightarrow S_S$ in that limit. With the aid of L'H\^opital's rule we get
\[2\lim_{\tau\rightarrow0}\left(S^{\tau}\frac{d\log_D S}{d\tau}\right)=\sumi p_i\left(\log_D p_i\right)^2-S_S^2>0.\]
The inequality follows from the fact that the right-hand side is the variance of $-\log_D p_i$. Hence, $d\log_D S/d\tau>0$ as $\tau\rightarrow0$, and so it is positive for all $\tau$. This implies that the logarithm of the exponential entropy (\ref{eq:S-exp}) is an increasing function of $\tau$, and that no stationary point given by condition (\ref{eq:stat}) exists.\par
This is easily confirmed from the expression for the R\'enyi entropy. Differentiating (\ref{eq:Renyi}) with respect to $\tau$ gives
\[\frac{dS_R}{d\tau}=-\frac{1}{\tau}\left[S_R+\frac{\sumi p_i^{1-\tau}\log_D p_i}{\sumi p_i^{1-\tau}}\right].\]
The stationary condition is again given by the escort average, (\ref{eq:stat}). The product $\tau d^2S_R/d\tau^2>0$ at the stationary point so that we have a local minimum for $\tau>0$ and a local maximum for $\tau<0$. This means that the curve of $S_R$ versus $\tau$ has a negative slope as it passes through $\tau=0$. The demonstration that $S_R$ has no extrema when considered as a function of $\tau$ follows exactly as before.
\par Hence, the R\'enyi entropy, or for that matter any mean \cite{Burrows}, cannot be expressed as an escort average because that would violate the condition that the mean is an increasing function of $\tau$. It is this property, in fact, which guarantees that the arithmetic mean ($\tau=1$) $\ge$ geometric mean ($\tau=0$) $\ge$ harmonic mean ($\tau=-1$). 
\section{Extent of a Distribution}
In the next to the last section we have found that the exponential of the R\'enyi entropy is the lower bound on the mean value of the number of different sequences of order $\tau$, whose equivalent mean was the mean of the HCDT entropy. Exponential mean entropies have been shown to be  measures of the extent of a distribution \cite{Campbell64}.\par
The measure of the extent of a distribution is inherently related to the error that is committed by using  an estimated probability distribution, $q$, when the \lq true\rq\ probability distribution is $p$. R\'enyi \cite{Renyi70} has referred to this as \lq information gain\rq, while Kullback \cite{Kullback}  used the term \lq directed divergence\rq, being based on the Shannon inequality
\begin{equation} \mathcal{E}_S(q|p)=\sumi p_i\log_D\left(\frac{p_i}{q_i}\right)
\ge0. \label{eq:E-S}
\end{equation}
The quantity $-\sumi p_i\log_D q_i$ is referred to as the inaccuracy \cite{Kerridge}.
Inequality (\ref{eq:E-S}) is easily seen to be a consequence of the arithmetic-geometric mean inequality: \begin{equation}
\prod_{i=1}^N\left(\frac{q_i}{p_i}\right)^{p_i}\le\sumi p_i\left(\frac{q_i}{p_i}\right)=1.\label{eq:ag}
\end{equation}\par
As a generalization of (\ref{eq:E-S}) we may consider the generating function 
\begin{equation}
\phi(q_i/p_i)=\frac{(q_i/p_i)^{\tau}-1}{\tau} \label{eq:T-bis}
\end{equation}
since the negative of its weighted arithmetic average in the limit as $\tau\rightarrow0$ is
\[-\lim_{\tau\rightarrow0}\sumi p_i\phi(q_i/p_i)=
\mathcal{E}_S(q|p).\]
The mean of (\ref{eq:T-bis}) in the same limit is
\[\lim_{\tau\rightarrow0}\phi^{-1}\left\{\sumi p_i\phi(q_i/p_i)\right\}=D^{-\mathcal{E}_S(q|p)}\le1,\]
which we will see to be related to the extent of a distribution.\par
The arithmetic average of (\ref{eq:T-bis}),
\[\sumi p_i\phi(q_i/p_i)=\frac{\sumi p_i^{1-\tau}q_i^{\tau}-1}{\tau},\]
has been referred to as the error incurred when the distribution $q$ is used instead of the \lq true\rq\ distribution, $p$ \cite[p. 208]{AD}. The mean value of (\ref{eq:T-bis}) has the equivalent mean 
\begin{equation}\mathfrak{M}_{\phi}(q/p)=\mathfrak{M}_{\tau}(q/p)=
\left(\sumi p_i^{1-\tau}q_i^{\tau}\right)^{1/\tau}\le1,
\label{eq:R-exp}
\end{equation}
since $\phi(x)$ is a linear function of $x^{\tau}$ for $\tau\neq0$ \cite[p. 68]{HLP}. The inequality in (\ref{eq:R-exp}) follows directly from H\"older's inequality,
\[\sumi p_i^{1-\tau}q_i^{\tau}\le\left(\sumi p_i\right)^{1-\tau}\left(\sumi q_i\right)^{\tau}\le 1,\]
for $0<\tau<1$. The second inequality makes allowance for incomplete distributions.\par The error  function with a parameter $\tau$,
\begin{equation}
\mathcal{E}_{R}(q|p)=-\frac{1}{\tau}\log_D\left(\sumi
p_i^{1-\tau}q_i^{\tau}\right)\ge0, \label{eq:E-R}
\end{equation}
 is constructed in analogy to the R\'enyi entropy \cite[p. 208]{AD}. In the $\tau=0$ limit, (\ref{eq:E-R}) reduces to the Shannon error function (\ref{eq:E-S}).\par
Because of the inequality of the means \cite[p. 26]{HLP}
\[\mathfrak{M}_r(x)\le\mathfrak{M}_s(x),\]
for $r<s$, with equality iff the probability distribution is uniform, the mean
\begin{equation}
\mathfrak{M}_{\tau}(q/p)=D^{-\mathcal{E}_R(q|p)} \label{eq:extent}
\end{equation}
provides a measure of the extent of a distribution. The mean (\ref{eq:extent}) is homogeneous in $q$, and satisfies
\begin{equation}
D^{-\mathcal{E}_S(q|p)}\le D^{-\mathcal{E}_R(q|p)}\le
\sumi q_i. \label{eq:extent-ineq}
\end{equation}
\par
The lower limit corresponds to the geometric mean 
\[
\mathfrak{M}_0(q/p)=\prod_{i=1}^N\left(\frac{q_i}{p_i}\right)^{p_i}=D^{-\mathcal{E}_S(q|p)}, \]
and is least affected by variations in $q/p$ \cite{Campbell64}, while the upper limit
\[\mathfrak{M}_1(q)=\sumi q_i \]
constitutes the \lq range\rq\ of the distribution. The range  is  most elementary measure of the extent of a distribution.\par
The difference between an ordinary and an \emph{incomplete\/} random variable is that the latter is not defined at every point in the sample space \cite[p. 570]{Renyi70}. Points where the random variable are undefined are said to be unobservable, and $\sumi q_i<1$ is the probability that the outcome will be observable.\par
The arithmetic-geometric mean inequality (\ref{eq:ag}) can be written as 
\begin{equation}\sumi q_i\ge D^{\sumi(p_i\log_D q_i-p_i\log_D p_i)},\label{eq:code-ineq}
\end{equation}
which is the first and second inequalities in (\ref{eq:extent-ineq}). Shannon's inequality (\ref{eq:E-S}) guarantees that the right-hand side of (\ref{eq:code-ineq}) is less than unity. The equality sign holds in (\ref{eq:code-ineq}) when $p_i=\lambda q_i$, where $\lambda=\left(\sumi q_i\right)^{-1}$, thus ensuring that $p$ is a complete distribution. This condition is obtained by requiring that $-\sumi p_i\log_D q_i$ be a minimum subject to $\sumi q_i=\mbox{const.}$
In fact, an essential part of the coding theorem for a noiseless channel is to show that the minimum of $-\sumi p_i\log q_i$ is $-\sumi p_i\log_D p_i$ subject to the constraint $\sumi q_i=1$ \cite[pp. 17--20]{Feinstein}. The latter constraint implies $\lambda=1$, and the Shannon error function vanishes leading to an equality in (\ref{eq:code-ineq}).\par However, there is another way the equality can be satisfied in  (\ref{eq:code-ineq}):  The minimum of $\sumi q_i$  subject to the constraint $\sumi p_i\log_D q_i=\mbox{const}$ is $D^{-\mathcal{E}_S(q|p)}$ \cite{Campbell64}. Introducing the Lagrange multiplier  $\lambda=q_i/p_i$, where $\lambda=\sumi q_i$, into the Shannon error function (\ref{eq:E-S}) results in \begin{equation}
\mathcal{E}_S(q|p)=\log_D\left(1\bigg/\sumi q_i\right).\label{eq:Boltzmann}
\end{equation}
An incomplete distribution $\sumi q_i<1$ leads to a finite error, $\mathcal{E}_S>0$.
The argument in the logarithm of (\ref{eq:Boltzmann}) can be interpreted as the number of digits necessary to specify the set of observable events. The smaller the set, the greater the number of digits that will be required.\par In this respect, (\ref{eq:Boltzmann}) is the antithesis of  Boltzmann's principle, where $\left(\sumi q_i\right)^{-1}$ is paired to the  \lq thermodynamic probability\rq, and the Shannon error function (\ref{eq:E-S}) to the entropy. Whereas Boltzmann's principle asserts that the greatest number of complexions, that correspond to a single macroscopic state, possesses the greatest entropy,  (\ref{eq:Boltzmann}) affirms that greatest number of digits needed to specify a given set, $\left(\inf\sumi q_i\right)^{-1}$,  corresponds to the greatest  error in discriminating between  two probability distributions. In other words, Boltzmann's principle is a measure of attenuation, whereas (\ref{eq:Boltzmann}) is a measure of accentuation.  \par
This is related to the problem of \lq\lq how to keep the forecaster honest\rq\rq \cite{Good}. A forecaster uses an estimated probability distribution $q$ to determine the outcome of events whose true distribution is $p$. His fee for correct prediction $f(q)$ is to be paid after it is known that the event has occurred. His expected fee is $\sumi p_if(q_i)$, which, if he is honest, must satisfy
\[\sumi p_if(q_i)\le\sumi p_if(p_i).\]
The Shannon error function (\ref{eq:E-S}) identifies $f$ with the logarithm. Then $\inf\sumi q_i$ in (\ref{eq:Boltzmann}) is that state with the lowest degree of predictability.\par
If  $q$ is the uniform distribution then the error (\ref{eq:E-R}) reduces to the differences in entropies so that the exponential of this difference is equal to the mean:
\begin{equation}
\mathfrak{M}_{\tau}\left(1/Np\right)=D^{S_R(p)-S_H\left(N\right)}. \label{eq:extent-bis}
\end{equation}
The closer $p$  is to the uniform distribution, the larger will be the mean value (\ref{eq:extent-bis}). Interpreting $\mathfrak{M}_{\tau}$ as a measure of extent, small values of $\mathfrak{M}_{\tau}$ imply that the probability measure is concentrated on a set of small $q$-measure for $0<\tau\le1$.\par As (\ref{eq:extent-bis}) shows, the difference between the Hartley and R\'enyi entropies is an exponential measure of the magnitude of $\mathfrak{M}_{\tau}$, and hence to the extent of the distribution. The fact that a small value of $\mathfrak{M}_{\tau}$ implies a small probability measure requires $\tau$ to be confined to the interval $(0,1]$ \cite{Campbell64} is thus related to the condition of concavity of the R\'enyi entropy. Only for $\tau\in(0,1]$ will the exponent in (\ref{eq:extent-bis})  be a true difference in entropies.\par

\end{document}